# Wrinkled few-layer graphene as highly efficient load bearer

Ch. Androulidakis[1], E.N. Koukaras[1], Jaroslava Rahova[2,3], Krishna Sampathkumar[2], John Parthenios[1], Konstantinos Papagelis[1,4], Otakar Frank[2] and Costas Galiotis[1,5]*

[1]Institute of Chemical Engineering Sciences, Foundation of Research and Technology-Hellas (FORTH/ICE-HT), Stadiou Street, Platani, Patras, 26504 Greece
[2]J.Heyrovsky Institute of Physical Chemistry of the CAS, v.v.i., Dolejskova 2155/3, 182 23 Prague 8, Czech Republic
[3]Institute of Geochemistry, Mineralogy and Mineral Resources,Faculty of Science, Charles University in Prague, Albertov 6, CZ 128 43 Prague 2, Czech Republic
[4]Department of Materials Science, University of Patras, Patras, 26504 Greece
[5]Department of Chemical Engineering, University of Patras, Patras 26504 Greece

*Corresponding author: c.galiotis@iceht.forth.gr or galiotis@chemeng.upatras.gr

## ABSTRACT

Multilayered graphitic materials are not suitable as load-bearers due to their inherent weak interlayer bonding (for example, graphite is a solid lubricant in certain applications). This situation is largely improved when two-dimensional (2-D) materials such as a monolayer (SLG) graphene are employed. The downside in these cases is the presence of thermally or mechanically induced wrinkles which are ubiquitous in 2-D materials. Here we set out to examine the effect of extensive large wavelength/ amplitude wrinkling on the stress transfer capabilities of exfoliated simply-supported graphene flakes. Contrary to common belief we present clear evidence that this type of "corrugation" enhances the load bearing capacity of few-layer graphene as compared to 'flat' specimens. This effect is the result of the significant increase of the graphene/polymer interfacial shear stress per increment of applied strain due to wrinkling and paves the way for designing affordable graphene composites with highly improved stress-transfer efficiency.

Graphene is the first isolated and most extensively studied nano-material among the 2-D family of layered crystals. It possesses remarkable mechanical properties, such as stiffness of 1 TPa, high extensibility and tensile strength as revealed by nano-indentation experiments on suspended flakes.[1] Graphene flakes consisting of more than one-layer in thickness have also been studied with the same method and the results showed that bi-layers and tri-layers have similar mechanical properties to the monolayer.[2] One of the most promising application of graphene is its use as a reinforcing agent in composite materials due to its extraordinary mechanical properties[3] and already commercial products are coming to the market. Despite the promising potential for mechanical enhancement of materials filled with graphene at low content,[4] there are still important issues that need to be overcome for exploiting the maximum potential of graphene.

One of these challenging issues is the presence of wrinkles in graphene structure which is a ubiquitous phenomenon and affects its mechanical behavior[5] and also the load-bearing capacity of the material. In single layer graphene, out-of-plane deformations due to thermal fluctuations are always present[6] and stress induced wrinkles are easily formed because of the very low bending rigidity of the material.[7] Suspended CVD mono-layers are thought as almost perfect graphene crystals,[8] their performance is compromised due to the extensive wrinkling they suffer as a result of the biaxial compression upon cooling from high temperatures[9-10] and the transfer process to other substrates.[11-12] As shown recently, a CVD monolayer resting or fully embedded in a PMMA/SU-8[13] or PET substrate[14] cannot be efficiently stressed because of the presence of large out-of-plane folds that reduce the stress transfer efficiency of the system.

Another important issue concerning the performance of few-layer (of two layers and more) graphene as reinforcing agent in composites, is the inherent stress transfer between the individual mono-layers that make up the few-layer flakes. In a polymer nano-composite material filled with few-layer graphene, the stress is transmitted to the graphene through shear forces developed at the interface of the outer mono-layers and the polymer. The stress is then transferred to the inner part of the layered material by shear through the weak van der Waals forces that bind the mono-layers. These weak secondary bonds are in fact responsible for the weak performance of all graphitic structures as load bearer materials[15], an example of which is the use of graphite as a solid lubricant in nuclear power stations. In the case of a bi-layer graphene, fully embedded in a matrix, both outer layers are directly adhered to the polymer and thus sufficiently stressed[14] but still the weak interlayer bonding affects somewhat the stress-transfer process. For a tri-layer or even thicker flakes, the middle layers easily slide in shear past each other and, therefore, the overall mechanical performance is compromised.[16] It is evident that for simply supported flakes on various substrates for which only one graphene surface is in contact with the substrate, the stress transfer efficiency of both bi-layer and tri-layer decrease significantly.[15]

One of the most efficient and widely used experimental techniques for studying the response of graphene under an external load is Raman spectroscopy[17-23] and is employed herein. The shift rates of the 2D and G Raman peaks with strain are well established and have been studied in various works under uniaxial and biaxial deformations.[18] Knowing the Raman phonon shift rates we can interrogate any graphene under an external stress and convert the Raman wavenumbers to values of strain (or stress in certain cases).[19] Thus, this methodology allows us to examine the stress take-up by the simply supported or embedded flakes[24] and to identify any modes of failure in tension or compression.[24]

In the present study we examine the effect of wrinkling in the mechanical performance of simply supported graphene with thickness ranging from one (1LG) to three (3LG) layers. Wrinkles with large wavelength/ amplitude ratios were thermally induced with a simple procedure. The wrinkled graphene flakes were examined under tension using Raman spectroscopy. In the case of mono-layer we found that that type of wrinkling did not affect its performance and the results were similar with previous studies. In the case of bi-layer and tri-layer we found that the induced wrinkling had a positive effect upon their tensile behavior and improved significantly their mechanical performance.

**Creation of wrinkled graphene**

Graphene flakes were produced by mechanical cleavage of graphite with the scotch tape method[25] and deposited directly on a PMMA/SU-8 polymer substrate. In order to create wrinkle patterns to the graphene flakes a biaxial stress field was generated by heating the polymer/graphene composite at 80º C for thirty minutes. Due to the thermal coefficient mismatch a biaxial stress field is generated and is relaxed by the creation of wrinkling patterns. We note that the substrate is also wrinkled along with the graphene as evident by both AFM and Raman measurements. In **figures 1a and c**, AFM images of an area with wrinkled flakes of varying thickness are presented and in the following **figures 1b and d,** the same area was scanned after the removal of the graphene. While the height channel shows the same wrinkles in both cases, the deformation channel clearly shows that (i) the wrinkles are confined only to areas under graphene (dark color in **figure 1c**), and (ii) the wrinkles remain imprinted into the substrate even after the graphene is removed. The deformation value (acquired using the PeakForce Quantitative Nanomechanical Property mapping mode of Bruker) expresses the depth into which the AFM tip is deforming the sample during contact at a given set-point.[26] Hence, the smaller the deformation, the stiffer the specimen. The yellow-brown (large deformation) bottom part of **figure 1c** is the softer polymer substrate, while the dark brown, upper, wrinkled part of the image is the graphene flake. In **figure 1d** the whole image is yellow-brown at the same color scale as **figure 1c**, reflecting the absence of graphene layers, regardless of their thickness. Only the polymer remained, though wrinkled, in the place where graphene was located. The removal of graphene is evidenced also by the Raman spectra (**figure 1e**), which show the presence of a bilayer graphene after the heating (black curve in e) and its absence after the sample is cleaned. We observe exactly the same wrinkling patterns on the surface of the polymer, which indicates clearly that the graphene/polymer is wrinkled as a unified system. This is a key feature as discussed below for the effective stress transfer from the polymer to the graphene flakes. Also, the wrinkles on the polymer are created only in the areas that were covered with graphene flakes, which show that the graphene drags with it the underline polymer while it relaxes the compressive stresses.

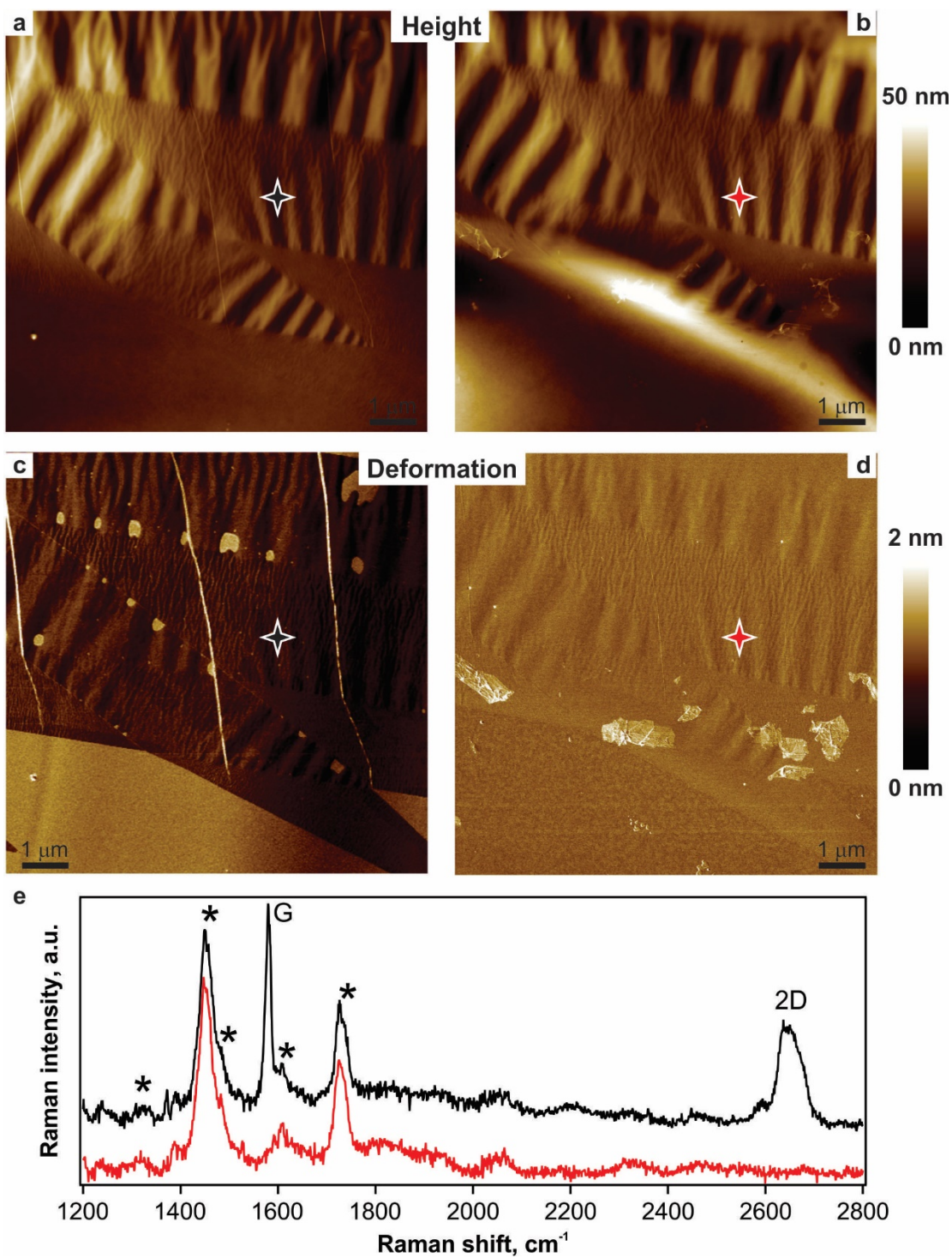

**Figure 1** AFM images (height sensor, a-b, and deformation, c-d) of samples with simply supported graphene flakes on SU8 after the heating (80°C) procedure (a,c) and after the removal of graphene (b,d) with a tape. Raman spectra (e) acquired at the spot indicated by the black and red star for samples with and without graphene, respectively (excitation wavelength, $\omega_{exc}$ = 633 nm). The stars in (e) denote the peaks of the PMMA.

**Figure 2** shows examples of SU8-supported graphene specimens with varying thickness, obtained by AFM. As can be seen from the AFM images, the wrinkle wavelength increases with the number of layers. Line profiles documenting the wrinkles evolution are depicted in **Figure 2d**. Even though the profiles show discontinuities and fluctuation both in amplitude and wavelength, most of the individual wrinkles have sinusoidal shapes. **Figure 2e** shows the medians of the wrinkle data (A, λ) for all measured samples. Most of the data points fall to a line with the slope of λ/A = 37.2.

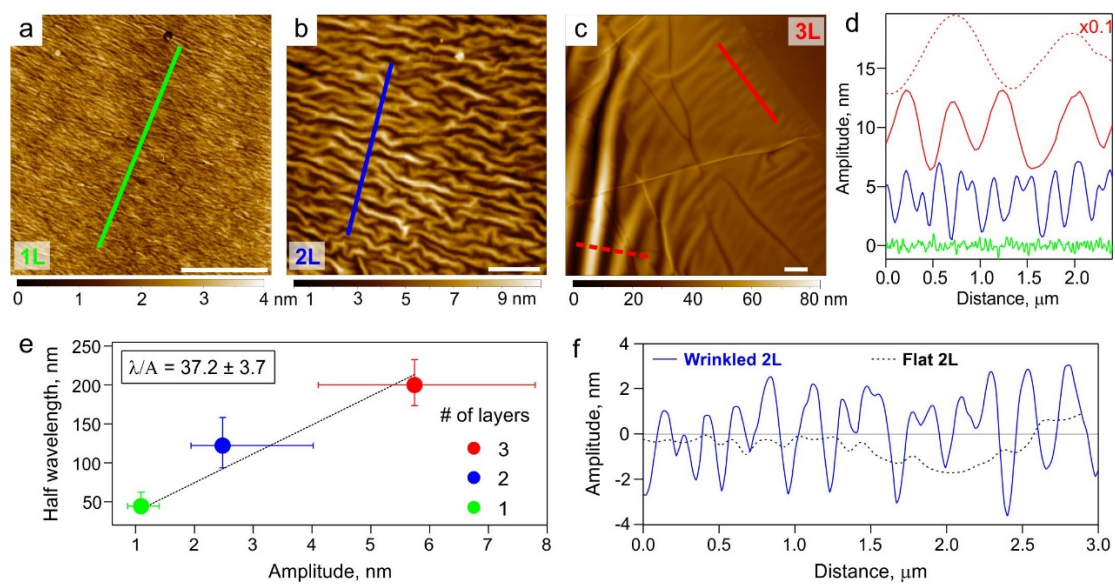

**Figure 2** (a-c) AFM images of simply supported graphene on SU8 with 1, 2 and 3 layers (d) Line profiles along the colored lines plotted in a-c. For ease of comparison, all profiles are cut to 2.5 μm; the curves are offset for clarity and the intensity of the top profile (dashed red line) is divided by 10. Line profiles along the colored lines plotted in a-c (green - 1L, blue - 2L, and red - 3L). (e) Median values of wrinkle parameters for simply supported graphene with 1-3 layers: half wavelength, λ, and amplitude, A. The error bars represent the interquartile range. The dashed line is the least squares line fit to the data, with the slope indicated in the black box; the error is the standard deviation. (f) Comparison of line profiles of a wrinkled and a flat bilayer.

In **figure S1** representative Raman spectra for the 2D peak of the examined graphene flakes are presented. The spectra were taken with a laser line of 785 nm. The examined flakes are mono-layer, and periodically stacked bi- and tri-layer. Optical images are also provided in the SI (**figure S2**).

**Mechanical response of wrinkled as compared to 'flat' graphene under tension**

The morphology of the examined wrinkled flakes as depicted from AFM measurements is presented in **figure 2.** Mono-layer (1LG), bi-layer (2LG) and tri-layer (3LG) wrinkled graphene samples were examined under tensile loading. From the projected area we calculate the residual compressive strain to be ∼−0.1% to −0.2%, in good agreement with the values of the position frequency of the 2D and G Raman peaks at the zero strain level[21] (**figure 3**). The graphene flakes contain wrinkles of high $\lambda/A$∼37 (**figure 2e**). For comparison 'flat' bi-layer and tri-layer flakes simply supported on polymer were also examined. The flat flakes were prepared by depositing exfoliated graphene flakes directly on cured polymer without any thermal treatment and the corresponding AFM images are presented in the SI (**figure S3**). For the 'flat' graphene, a very small fluctuation across the flakes is still present with amplitude of ∼ 0.4 nm (**figure 2f**), which is one order of magnitude smaller than the wrinkle amplitude induced thermally. Thus, these flakes can be considered as practically flat with a small 'rippling' effect and their morphology is clearly distinguished from the wrinkled specimens in which smooth periodic wrinkling patterns are observed (**figure 2f**). The samples were subjected to incremental tension using a four-point-bending jig and Raman spectra (2D and G peaks) were recorded *in situ* at elevated strain levels. More details can be found in the experimental section. This is now a well-established technique for testing two dimensional crystals under axial mechanical loadings combined with Raman spectroscopy[17-19, 24].

In **figure 3a,b** the position of the Raman 2D and G peaks for the wrinkled mono-layer graphene versus the applied tensile strain is presented. The shift rate for the 2D peak is ∼ −57.4 cm$^{-1}$/% in excellent agreement with previously reported values[17, 19] for pristine ('flat') flakes. The G mode clearly splits into the G$^-$ and G$^+$ peaks, with shift rates of −29.6 cm$^{-1}$/% and −9.7 cm$^{-1}$/% (**figure 3b**) confirming that the wrinkled mono-layer has been effectively stressed. The experimental results of the present work are summarized in **Table 1**. The value of the shift rate is the average of the response from fifteen points located at distances of a few microns from each other, covering a large area of the examined graphene over 100 μm$^2$. This indicates that the shift values are representative of the whole specimen and cannot be attributed to random phenomena or local abnormalities.

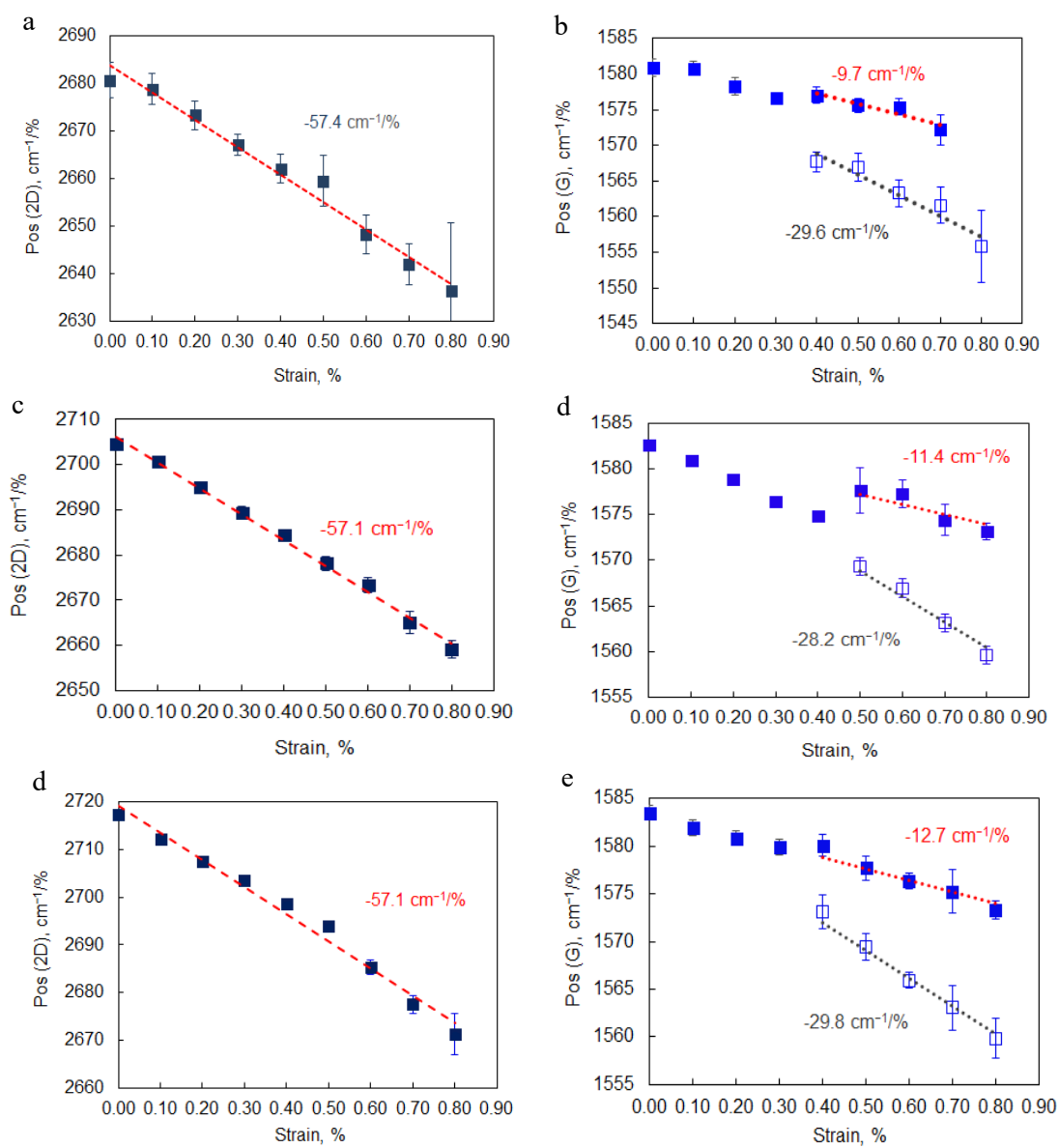

**Figure 3** The shift of the position frequency of the 2D and G Raman peaks under tensile strain for wrinkled (a), (b) mono-layer, (c), (d) bi-layer and (e), (f) tri-layer graphene flakes simply supported on polymer ($\omega_{exc}$ = 514nm). In all cases the flakes are ideally stressed.

In the case of CVD graphene, the wrinkles are the reason for the lower shift rates reported in other studies[14, 27] (and consequently for their lower reinforcing potential in nano-composites), while the present wrinkling characteristics have no influence on mechanical response of the exfoliated mono-layer.

**Table 1** The shift rates for simply supported graphene with thickness of one to three layers under tension for the 2D and G Raman peaks. The error represents the standard deviation of the linear fit.

|  | Thickness (No of layers) | $\partial Pos(2D)/\partial\varepsilon$ (cm$^{-1}$/%) | $\partial Pos(G^+)/\partial\varepsilon$ (cm$^{-1}$/%) | $\partial Pos(G^-)/\partial\varepsilon$ (cm$^{-1}$/%) |
|---|---|---|---|---|
| Wrinkled flakes | 1 | −57.4 ± 3.0 | −9.7 ± 0.8 | −29.6 ± 1.1 |
|  | 2 | −57.1 ± 1.5 | −11.4 ± 1.5 | −28.2 ± 1.3 |
|  | 3 | −57.1 ± 2.8 | −12.7 ± 1.1 | −29.8 ± 1.0 |
| «Flat» flakes | 2 | −48.6 ± 2.5 | −11.6 ± 0.7 | −27.9 ± 1.0 |
|  | 3 | −28.6 ± 1.4 | −6.5 ± 1.8 | −18.3 ± 1.2 |

The experimental results for the wrinkled simply supported bi-layer graphene for the 2D peak are presented in **figure 3c**. A line with length of 20 microns near the geometric center of the flake and parallel to the direction of the tension was mapped across the bi-layer with step of 1 μm. The shift rate of the 2D Raman peak is ∼ −57.1 cm$^{-1}$/%, and −28.2 cm$^{-1}$/% and −11.4 cm$^{-1}$/% for the shift rates of G$^-$ and G$^+$, respectively, averaged from all measured points. The results for the 'flat' bi-layer are presented in **figure 4a,b** and the shift rates in this case are ∼ −48.6 cm$^{-1}$/% for the 2D peak (ignoring the slack in the initial strain levels) and −27.9 cm$^{-1}$/% and −11.6 cm$^{-1}$/% for the G$^-$ and G$^+$, respectively, averaged from all measured points. The values for the 'flat' bilayer are higher compared to previous studies[17] for the same case and at similar level with the wrinkled graphene. This effect is more pronounced for the trilayer flakes for which theoretical simulations have documented that by applying uniaxial tension of 1% on the top and bottom layer the middle layer deforms at about 0.1% implying that interlayer coupling cannot transfer the external load efficiently[28]. As explained in detail later, the observed enhanced performance is attributed to the low-wavenumber wrinkled morphology. The two examined cases are presented schematically in **figure 5b,d**.

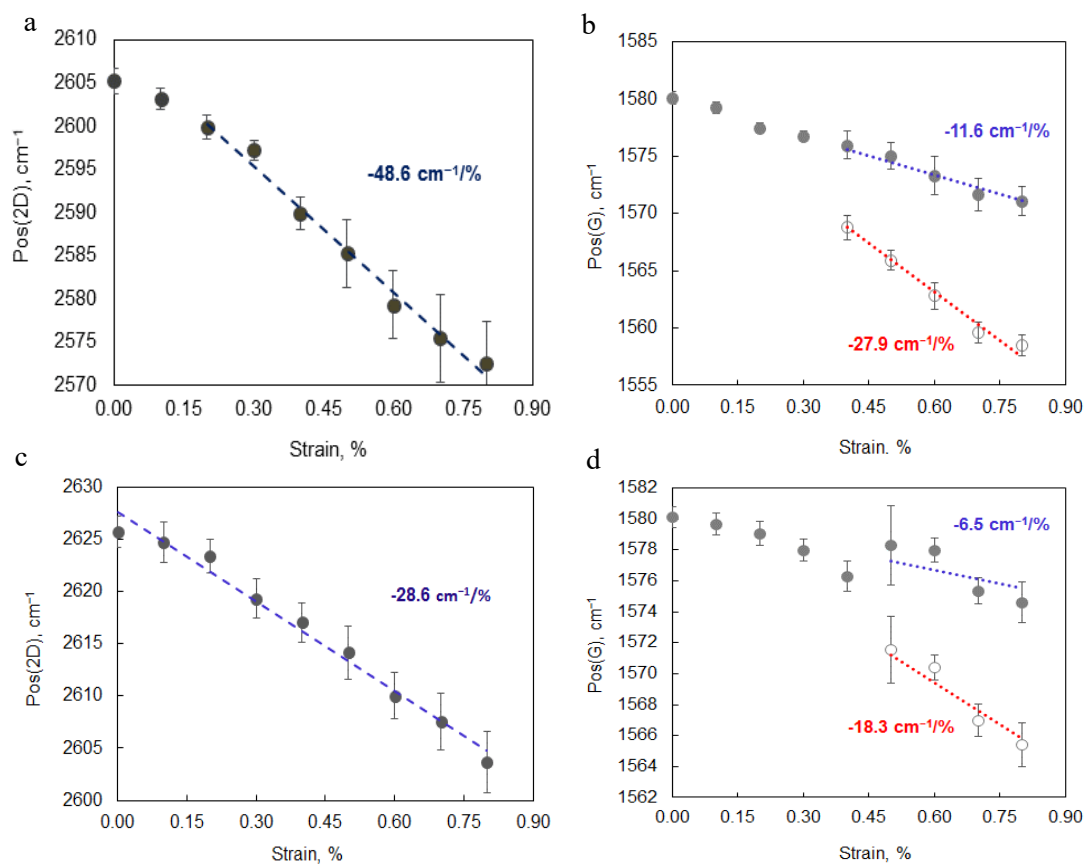

**Figure 4** The shift of the position of the 2D and G Raman peaks under tensile strain for 'flat' (a), (b) bi-layer, (c), (d) tri-layer graphene flakes simply supported on polymer ($\omega_{exc}$ = 514 nm).

The morphology of the examined wrinkled tri-layer is depicted in **figure 1c** and in **figure 3e** the shift of the 2D Raman peak with tension for the wrinkled tri-layer is shown. The value of the shift rate is ~ −57.1 cm$^{-1}$/% averaged from 10 points measured across a line with length of 20μm. The shift value is in agreement with the results obtained from the wrinkled mono-layer and bi-layer. Additional results from other tri-layer flake with similar response are presented in SI (**figure S5**). **Figure 4c** shows the corresponding results for the 'flat' tri-layer. The shift rate is considerably lower with its value being about a half of the wrinkled one. The shift rate of the wrinkled tri-layer is larger than uncoated tri-layer examined previously[15] and even higher than embedded tri-layer[15-16, 29].

The shift rates for the two G sub-peaks in the wrinkled tri-layer are −29.8 cm$^{-1}$/% and −12.7 cm$^{-1}$/% for the G$^-$ and G$^+$, respectively, averaged from all measured points (**figure 3f**). A substantial decrease is detected in the shift of the G$^-$ and G$^+$ for the 'flat' flake, too. Comparing to previous works, the G peak shift rates of the wrinkled tri-layer are higher than previously reported results for an embedded[15-16] tri-layer. The wrinkled tri-layer also achieves its optimal performance despite that the normal stress is transmitted to the flake via shear of only the bottom layer which adheres to the polymer. These results are reproducible from a whole range of large 2LG and 3LG flakes and thus, the response should be considered representative for the examined flakes.

The experimental results show that a specific wrinkling network with high wavelength over amplitude ratio ($\lambda/A$) has in fact a positive effect upon the mechanical performance of simply-supported graphene. The effect of wrinkling on the stress transfer process has also been examined recently in monolayer CVD graphene under tension and shift rates significantly lower[14, 27] than the value of $-60$ cm$^{-1}$/% were obtained. However, the wrinkling morphology in the cases of CVD graphene is vastly different than the corrugation morphology induced here. As it happens, in the CVD production procedure, compression of magnitude 2-3% is developed (upon cooling down from 1000°C)[30], and is subsequently relaxed upon cooling by the creation of wrinkles and folds (not in contact with the substrate) with amplitude of up to 20 nm.[30] This type of morphology is also present in graphene on the target substrate after the transfer process and brings about mechanically isolated islands. The diameter of these islands is smaller than the transfer length required for full stress transfer as required by the shear lag principles,[14] and as a result only a fraction of the applied stress is transferred from the substrate to the graphene.[15] Even though the dimensions of those islands can be modified to some extent depending on the transfer method,[13] the delaminated wrinkles which cause a sincere drop in the interfacial stress transfer are present on different kinds of polymers as PET[14], PMMA as well as on SU-8.[13] On the other hand, in the present case, the graphene conforms well on the substrate without the presence of delaminated areas. This can be deduced by the fact that the polymer is also wrinkled along with the graphene as a unified system (**figure 1**).

Previous MD simulations have shown that small deviations from the ideal bonding even at the angstrom scale can affect the mechanical behavior of graphene on polymers by reducing the adhesion energy.[31-32] As mentioned earlier, this is exactly the case for many systems particularly those employing

CVD graphene, since in those cases the 2-D inclusion is already heavily wrinkled exhibiting out-of-plane folds which are detrimental as far as stress-transfer is concerned.[33] This case is clearly depicted in **figure 5c** bearing in mind that the size of the islands in-between adjacent folds is below the critical length required for efficient load transfer. In the case examined here, the corrugated morphology and its perfect attachment to the underlying substrate (**figure 5d, e**) results in the increase of the interfacial shear between the two materials and brings about a more efficient built-up of normal stress.

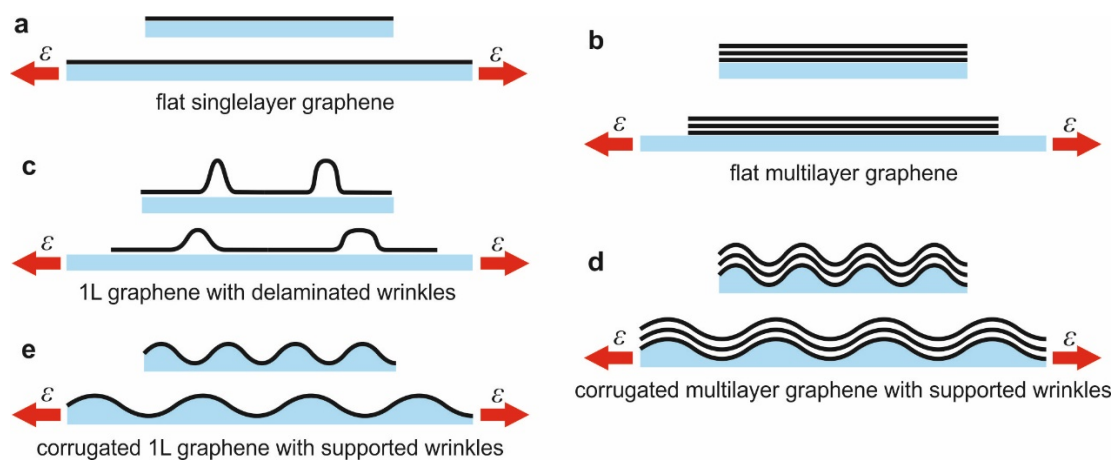

**Figure 5** Schematic of the morphology and qualitative stress transfer of a flat mono- (a) and few-layer (b), of a wrinkled CVD mono-layer on polymer (c), and wrinkled few- (d) and mono-layer (e) graphene.

The optimal mechanical response of the few-layer wrinkled graphene can be simply explained by the increase of the interfacial shear stress per increment of applied load due to the wrinkle morphology. In order to verify this, we examined the stress transfer mechanism in a wrinkled tri-layer/polymer system under tension by collecting Raman maps across its length for various strain levels. A corresponding AFM image of the examined tri-layer is provided in SI (**figure S4**). The length of the flake is about ~20 μm. Line profiles of the Raman 2D peak frequency for various strain levels are presented in **figure 6a**. The stress transfer mechanism is, in fact, similar to that reported for the case of simply supported mono-layer.[21, 34-35] We can estimate the shear stress at the interface of the wrinkled tri-layer graphene/polymer per increment of applied strain at room temperature by the balance of shear and axial forces:[21]

$$\left(\frac{\partial \varepsilon}{\partial x}\right)_{T \equiv 298K} = -\frac{\tau_t}{nt_g E} \Leftrightarrow \tau_t = -nt_g E \left(\frac{\partial \varepsilon}{\partial x}\right)_{T \equiv 298K} \qquad (1)$$

Where $\varepsilon$ is the strain, $\tau_t$ is the interfacial shear stress, $E$ is the Young's modulus of graphene, $n$ is the number of layers of the graphene and $t_g$ is the thickness of a single layer graphene. As mentioned earlier the slopes $d\varepsilon/dx$ (**figure 6a**) can be extracted from the Raman data and therefore the values of the interfacial shear stress per strain level can be easily estimated (**figure 6b).** Detailed analysis for this experimental procedure is provided in the SI. The maximum shear stress obtained at a maximum tensile strain level of ~0.80% is estimated to be ~0.75 MPa (see SI). This value is much higher than values obtained from 'flat' –simply supported-graphene/ polymer systems[21, 34, 36-37] for similar strain levels. Interestingly enough, the shear stress for the wrinkled tri-layer obtained herein is more than twice higher than the corresponding $\tau_t$ reported for a bi-layer/polymer system.[15] Moreover, the MD simulations performed on a mono-layer graphene/polymer showed that the interfacial shear strength is also higher for a wrinkled mono-layer graphene compared to a flat one,[38] which further supports the present findings. It is also expected that the wrinkling network also leads to higher shear between the individual single graphene layers due to enhanced interlayer friction, since previous theoretical studies also showed that wrinkling increases the friction significantly compared to flat flakes.[39-42]

     The above results suggest that the interfacial shear stress of graphene/polymer per applied uniaxial strain can be tuned by the creation of such wrinkle patterns. Another method to increase the generated interfacial shear stress is by chemically altering the interface of graphene/polymer by oxidative or similar treatments.[43] Our suggestion is more advantageous than chemical surface treatments as no defects are induced and therefore the mechanical integrity of the inclusion is not impaired. Another important factor is the transfer length for efficient stress-transfer which has not been considered in past attempts on 'flat' flakes[15]. As is evident from **figure 6a** in the wrinkled case one needs a flake of length ≥ 20 μm in order to get efficient loading at the geometric middle of the flake. One assumes that for the 'flat' for which the

interface is weaker much larger transfer lengths are required and this can partly explain the low values of Raman wavenumber shifts reported previously[15].

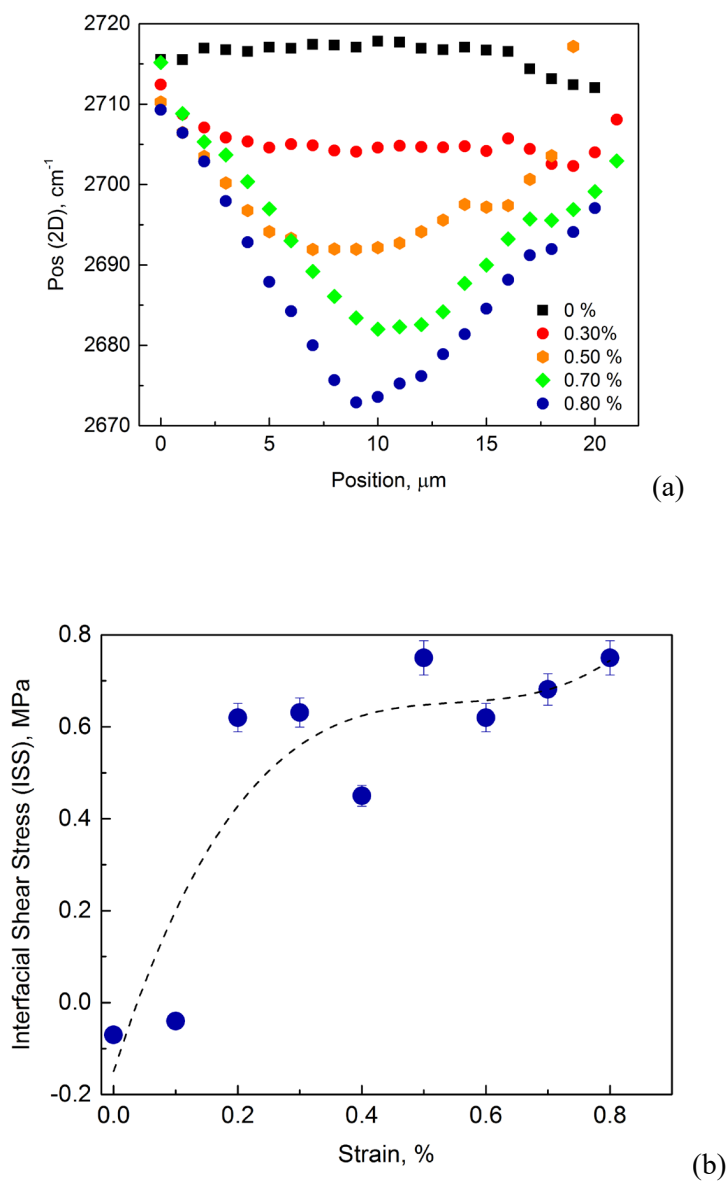

**Figure 6** (a) Raman shift of the 2D peak across the length of a wrinkled tri-layer graphene and (b) interfacial shear stress of graphene/polymer system for various strain levels of tension (the dashed line is a guide to the eye).

Another advantage for using high wavelength over amplitude wrinkling in few-layer graphene is the fact that the modulus of ~1 TPa is still retained since the maximum Raman shifts per strain (~57 cm$^{-1}$/%) were obtained for all cases[44] (see above). The second advantage concerns the volume fraction when graphitic materials are used. It is obvious that the bi-layer and tri-layer will provide higher volume fraction[15] compared to the mono-layer with the same dimensions, respectively, while there is no trade-off between the effective modulus and the thickness of graphene. Thus, there is no need to resort in the exclusive use of mono-layers, which are more difficult and expensive to produce. The only downside is the relatively larger transfer length that is required for full stress transfer from the polymer to the graphene, although this is compensated by the fact that it is easier to produce thicker graphene with larger length.[45] Further investigation of all these factors is currently under way.

This type of wrinkling is useful and for other aspects of composite materials where the graphene flakes are fully embedded in polymers. Recently, graphite was exfoliated directly in an epoxy resin without any additives and the filling graphene flakes had thickness of a few nano-meters[46]. The effect of temperature on the exfoliation was also examined and very good enhancement to the mechanical and electrical properties was observed in the resulting composite material. Because of the similarity in the exfoliation process (exfoliation using heat treatment and directly on polymer), our results suggest that the wrinkled flakes with conformal adhesion to the polymer can potentially provide a remarkable enhancement not only in the mechanical properties, but in the electrical and thermal properties as well, even with tri-layer graphene flakes.

In summary, we examined wrinkled/ 'flat' graphene on a polymer substrate under tensile deformation with thickness of one to three layers. We found that the wrinkling network does not compromise the mechanical response of mono-layer graphene and its performance is in broad agreement with previous studies.[17, 19, 27] Concerning the bi-layer and tri-layer flakes, we found that the wrinkled flakes present similar shift rates to the mono-layer which are significantly higher than previous results obtained from 'flat' flakes. Henceforth, this small wavelength wrinkling seems to assist few layer graphene to achieve optimal performance by increasing the stress transfer due to the enhanced shear strength of the graphene/polymer interface. This is an important finding because it can turn the unavoidable formation of wrinkles during the production procedures to our advantage and also to allow the use of thicker flakes of same level of performance as the mono-layer. Practical application of the present results can lead to the design of nano-composite materials with relatively higher graphene volume fractions and henceforth superior mechanical properties.

**Methods**

Graphene flakes were prepared by mechanical exfoliation of graphite using the scotch tape method and deposited directly on a PMMA/SU-8 substrate. The SU-8 (SU-8 2000.5 Micro- Chem) is a photoresist polymer and was spin coated in the top surface of a PMMA bar at ~ 2000 rpm creating a layer of about ~300 nm thickness. The sample polymer/graphenes was heated at 80°C for thirty minutes in order to induce compressive stresses. Using this method, a residual compressive strain of ~ −0.1% to −0.2% was observed for all samples. We note that the SU-8 was not cured before the deposition of the graphitic materials, in order to be compliant for the creation of wrinkles in the graphenes and avoid local delamination. The samples were characterized by means of AFM (see below) and Raman spectroscopy, either by LabRAM HR (Horiba) microspectrometer using 633 nm laser excitation, and 600 groves/mm grating, or InVia (Renishaw) microspectrometer with 2400 groves/mm grating for the 514 nm laser excitation; 100x long working distance objective was used in all cases.

The tensile experiments were conducted using a four-point-bending jig, designed to adjust under a Raman microscope for simultaneously applying strain and collecting Raman spectra. A laser line of 514.5 nm was used in the case of wrinkled graphenes and a laser line of 785 nm for the flat flakes. The increment step of strain was 0.1 % for all cases.

The AFM images were obtained using Dimension Icon or FastScan microscopes (Bruker) operating in Peak Force Tapping mode using ScanAsyst-Air probes (stiffness 0.2-0.8 N/m, frequency ~80 kHz). No data treatment apart from line subtraction (retrace) to remove the tilt has been performed. Wrinkle amplitudes (A) and half wavelengths (λ) were measured from line profiles manually for each wrinkle. Medians of the thus obtained parameters were then used to characterize the wrinkles in each profile. If possible, more (>3) profiles were measured for one flake, in that case the average value of the profile medians with one standard deviation (both for the amplitude and the wavelength) was used. The 'AFM compression' value was obtained as a ratio between the horizontal and surface distances for a particular profile.

To extract the dε/dx slopes from the Raman data, the difference of the Raman shifts were taken with respect to a reference value of 2710.34 cm-1, that corresponds to a zero strain level. The resulting values were then converted to strain values using the strain rate, $k_{2D}$, of the 2D Raman band. For this we used the representative shift per percent of strain discussed above, obtained by the wrinkled bi- and tri-layer flakes, of −57.1 cm-1/%. The resulting datasets were then smoothed using a Savitzky-Golay filter[47] with a quadratic polynomial and a 7-11point window size (depending on the irregularities/noise of the datasets). The interfacial shear stress (ISS) corresponds to the first derivative of the smoothed dataset with the proper multiplication factors (see eqn. 1 in the main text).

*Supporting Information Available*: Raman spectra of the 2D peak for the examined flakes of various thicknesses and the corresponding optical images. Optical image and AFM topography of the 'flat' flakes. Details on the stress transfer mechanism for the wrinkled trilayer. Topography and shift rate of the 2D peak with tension. Detailed Raman maps for all strain levels measured along with associated fittings. This material is available free of charge via the Internet at http://pubs.acs.org.

**Acknowledgements**

The authors acknowledge the financial support of the Graphene FET Flagship (''Graphene-Based Revolutions in ICT And Beyond''- Grant agreement no: 604391) and of the European Research Council (ERC Advanced Grant 2013) via project no. 321124, "Tailor Graphene". O.F., J.R. and K.S. acknowledge the support of Czech Science Foundation project No. 14-15357S.

The authors declare no competing financial interest.

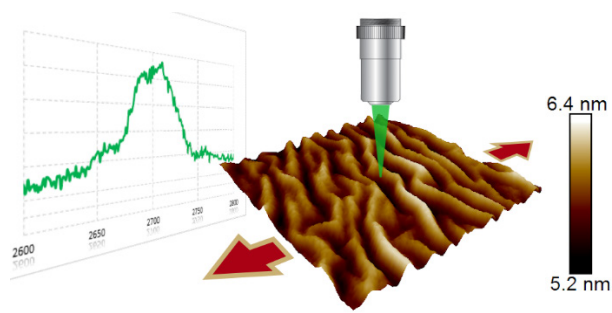